\documentclass[reprint,noshowpacs,noshowkeys,prd,balancelastpage,nofootinbib]{revtex4}
\usepackage[utf8]{inputenc}
\usepackage{color}
\usepackage{mathrsfs}
\usepackage{amsmath}
\usepackage{amssymb}
\usepackage{graphicx}
\usepackage[unicode=true,
 bookmarks=false,
 breaklinks=false,pdfborder={0 0 1},backref=section,colorlinks=true]
 {hyperref}
\hypersetup{
 linkcolor=blue,urlcolor=blue,citecolor=red}

\makeatletter
\@ifundefined{textcolor}{}
{%
 \definecolor{BLACK}{gray}{0}
 \definecolor{WHITE}{gray}{1}
 \definecolor{RED}{rgb}{1,0,0}
 \definecolor{GREEN}{rgb}{0,1,0}
 \definecolor{BLUE}{rgb}{0,0,1}
 \definecolor{CYAN}{cmyk}{1,0,0,0}
 \definecolor{MAGENTA}{cmyk}{0,1,0,0}
 \definecolor{YELLOW}{cmyk}{0,0,1,0}
}

\usepackage{amsfonts}
\usepackage{mdframed}
\usepackage{footnote}
\usepackage{graphicx}
\usepackage{float}
\usepackage[font=footnotesize,it]{caption}
\usepackage{tikz}
\usepackage{tikz-3dplot}

\makeatother

\begin{document}
\title{$\mathcal{PT}$-Symmetric Generalized Extended Momentum Operator }
\author{M. Izadparast }
\email{masoumeh.izadparast@emu.edu.tr}

\author{S. Habib Mazharimousavi}
\email{habib.mazhari@emu.edu.tr}

\affiliation{Department of Physics, Faculty of Arts and Sciences, Eastern Mediterranean
University, Famagusta, North Cyprus via Mersin 10, Turkey}
\date{\today }
\begin{abstract}
We develop further the concept of generalized extended momentum operator
(GEMO), which has been introduced very recently in \citep{M.H2},
and propose the so called $\mathcal{PT}$-symmetric GEMO. In analogy
with GEMO, the $\mathcal{PT}$-symmetric GEMO also satisfies the extended
uncertainty principle (EUP) relation. Moreover, the corresponding
Hamiltonian that is constructed upon the $\mathcal{PT}$-symmetric
GEMO, with a real or $\mathcal{PT}$-symmetric potential, remains
non-Hermitian but $\mathcal{PT}$-symmetric and consequently its energy
and momentum eigenvalues are real. We apply our formalism to a quasi-free
quantum particle and the exact solutions for the energy spectrum are
presented. 
\end{abstract}
\maketitle

\section{Introduction}

The $\mathcal{PT}$-symmetric quantum mechanics has been introduced
by Bender and Boettcher in 1998 \citep{BenderBoettcher1998,Complex correspondence principle,Making sense,Potentials}.
In spite of Hermiticity which is a pure mathematical principle, $\mathcal{PT}$-symmetry
is a physical axiom. It is a form of non-Hermitian operator representing
the parity ($\mathcal{P}$) and time reversal ($\mathcal{T}$) symmetry
in a quantum system. It is able to define a Hamiltonian to be an observable
due to producing real energy eigenvalues. Additionally, the probability
is conserved holding the time evolution unitary. In a quantum system,
the momentum is not only a Hermitian operator, but also it keeps the
$\mathcal{PT}$-symmetry attitudes. To find a $\mathcal{PT}$-symmetric
Hamiltonian, one has to examine the potential, $V(x)$, whether it
meets the $\mathcal{PT}$-symmetric criteria or not. In this respect,
the potential approves the parity and time reversal symmetries i.e.,
$\mathcal{PT}V\left(x\right)\equiv V^{\star}(-x)=V\left(x\right)$.
There exist several remarkable studies in the sense of non-Hermitian
Hamiltonian concerning $\mathcal{PT}$-symmetric potential in the
literature \citep{Potentials}. Also, the transition between broken
and unbroken $\mathcal{PT}$-symmetry regions has been observed experimentally
\citep{Experiments}. Moreover, Mostafazadeh has given a wider view
of non-Hermitian quantum theory in a series of works entitled as pseudo-Hermitian
\citep{Mostafazadeh}. 

Keeping in mind the value of the short history of $\mathcal{PT}$-symmetric
Hamiltonians, now in this study, we introduce an alternative approach
for producing $\mathcal{PT}$-symmetric Hamiltonians using $\mathcal{PT}$-symmetric
GEMO. The latter contribution of the novel form of momentum is incorporated
with the extended uncertainty principle (EUP) which is discussed earlier
in \citep{M.H2}. We propose a complexified GEMO (CGEMO) consist of
a $\mathcal{PT}$-symmetric auxiliary function $\mu(x)$ such that
the CGEMO satisfies the so called extended uncertainty principle (EUP)
commutation relation. Furthermore, we construct the corresponding
$\mathcal{PT}$-symmetric Hamiltonian, utilizing the CGEMO. On the
other hand, since the Hamiltonian and $\mathcal{PT}$operator commute,
i.e., $[\mathcal{H},\mathcal{PT}]=0$, the $\mathcal{PT}$-symmetry
of such a system remains unbroken.

This paper is organized in the following order. In Sec. II, we propose
the concept of CGEMO and prove that it is an observable in the sense
that its eigenvalues are real. Afterwards, in Sec. III, one example
of complex GEMO is demonstrated with the corresponding eigenfunctions
and energy eigenvalues. In Conclusion we summarize our paper.

\section{Observable Generalized Momentum operator}

Recently we have studied the Generalized Extended Momentum Operator
(GEMO) \citep{M.H2} in the context of EUP, given by

\begin{equation}
p=-i\hbar\left(1+\mu\left(x\right)\right)\frac{d}{dx}-\frac{i\hslash}{2}\frac{d\mu\left(x\right)}{dx},\label{eq:1}
\end{equation}
 in which the auxiliary function $\mu\left(x\right)$ is a real function
of position operator $x$. The GEMO is Hermitian, $p=p^{\dagger}$
and satisfies the EUP relation, i.e., 
\begin{equation}
\left[x,p\right]=i\hbar\left(1+\mu\left(x\right)\right).\label{GEMO1}
\end{equation}
Here, in this study we propose $\mu\left(x\right)$ to be a $\mathcal{PT}$-symmetric
complex function i.e., $\mathcal{PT}\mu\left(x\right)\equiv\mu^{\star}\left(-x\right)=\mu\left(x\right)$
which in turn makes the GEMO non-Hermitian but $\mathcal{PT}$-symmetric.
This means that, $p\neq p^{\dagger},$but $\left[\mathcal{PT},p\right]=0$.
Having GEMO to be $\mathcal{PT}$-symmetric, implies that its eigenvalues
are real. To prove that we start from the eigenvalue equation of the
momentum operator 
\begin{equation}
p\Phi_{\mathfrak{p}}\left(x\right)=\mathscr{\mathit{\mathfrak{p}}}\Phi_{\mathfrak{p}}\left(x\right)\label{eq:5}
\end{equation}
 in which $\mathfrak{p}$ and $\Phi_{\mathfrak{p}}$ are the momentum
eigenvalue and corresponding eigenfunction, respectively. Applying
$\mathcal{PT}$operator from the left side on (\ref{eq:5}) yields
\begin{equation}
\mathcal{PT}\left(p\Phi_{\mathfrak{p}}\left(x\right)\right)=\mathfrak{p}^{\star}\left(\mathcal{PT}\Phi_{\mathfrak{p}}\left(x\right)\right)\label{eq:6}
\end{equation}
which after the fact that $p$ and $\mathcal{PT}$ commute, one finds
\begin{equation}
\mathfrak{p}\Phi_{\mathfrak{p}}\left(x\right)=\mathfrak{p}^{\star}\Phi_{\mathfrak{p}}\left(x\right)\label{eq:7}
\end{equation}
which in turn implies that $\mathfrak{p}=\mathfrak{p}^{\star}$, indicating
$\mathfrak{p}$ is real and consequently, $p$ is a physical observable. 

Furthermore, the corresponding Hamiltonian of a particle with a $\mathcal{PT}$-symmetric
GEMO undergoing a one-dimensional potential $V\left(x\right)$ is
given by \citep{M.H2}
\begin{equation}
\mathcal{H}=\frac{-\hbar^{2}}{2m}\left(\left(1+\mu\right)^{2}\frac{d^{2}}{dx^{2}}+2\left(1+\mu\right)\mu^{\prime}\frac{d}{dx}+\frac{1}{2}\left(1+\mu\right)\mu^{\prime\prime}+\frac{1}{4}\left(\mu^{\prime}\right)^{2}\right)+V\left(x\right)\label{eq:Hamil}
\end{equation}
in which a prime stands for the derivative with respect to $x$. Herein,
with a $\mathcal{PT}$-symmetric potential i.e., $\mathcal{PT}V\left(x\right)=V\left(x\right)$,
$\mathcal{H}$ becomes $\mathcal{PT}$-symmetric which indicates 
\begin{equation}
\left[\mathcal{PT},\mathcal{H}\right]=0.
\end{equation}
In a similar manner as of GEMO, one can show that the eigenvalues
of the Hamiltonian are real, i.e., in 
\begin{equation}
\mathcal{H}\phi\left(x\right)=E\phi\left(x\right)\label{SE}
\end{equation}
$E$ is real. Next, we apply the so-called Point-Canonical-Transformation
(PCT), upon which we introduce
\begin{equation}
z=z\left(x\right)=\int^{x}\frac{1}{1+\mu\left(y\right)}dy+z_{0}\label{GEMO2}
\end{equation}
in which $z_{0}$ is a gauge constant and the wavefunction is decomposed
as
\begin{equation}
\phi\left(x\right)=\frac{1}{\sqrt{1+\mu\left(x\right)}}\chi\left(z\left(x\right)\right)\label{GEMO3}
\end{equation}
to transformed the Schrodinger equation (\ref{SE}) to the standard
form of the Schrodinger equation in $z$-space given by
\begin{equation}
\frac{-\hbar^{2}}{2m}\frac{d^{2}}{dz^{2}}\chi\left(z\right)+V\left(x\left(z\right)\right)\chi\left(z\right)=E\chi\left(z\right).\label{GEMO4}
\end{equation}
We observe here that, due to the PCT, although the Schrodinger equation
in $x$-space is transformed into the standard Schrodinger equation
in $z$-space, but the energy eigenvalues remained the same. In the
other words, the energy is invariant under the PCT transformation. 

\section{Quasi free particle\textmd{\textup{\normalsize{} }}}

Following the first example in Ref. \citep{M.H2}, we choose 
\begin{equation}
\mu\left(x\right)=\alpha^{2}x^{2}+i2\beta x\label{eq:Au1}
\end{equation}
 in which $\alpha$ and $\beta$ are two real constants such that
$\mu\left(x\right)$ remains $\mathcal{PT}$-symmetric. After, (\ref{eq:Au1}),
the CGEMO becomes 

\begin{equation}
p=-i\hbar\left(1+\alpha^{2}x^{2}+i2\beta x\right)\frac{d}{dx}-i\hslash\left(\alpha^{2}x+i\beta\right)\label{eq:8}
\end{equation}
which is $\mathcal{PT}$-symmetric and satisfies the EUP relation
$[x,p]=i\hbar\left(1+\alpha^{2}x^{2}+2i\beta x\right)$. Here let's
find the eigenfunctions and eigenvalues of the CGEMO. To do so we
start with
\begin{equation}
p\Phi_{\mathfrak{p}}\left(x\right)=\mathscr{\mathit{\mathfrak{p}}}\Phi_{\mathfrak{p}}\left(x\right)
\end{equation}
which yields
\begin{equation}
\left(1+\alpha^{2}x^{2}+i2\beta x\right)\frac{d\Phi_{\mathfrak{p}}\left(x\right)}{dx}+\left(\alpha^{2}x+i\beta-\frac{i\mathfrak{p}}{\hslash}\right)\Phi_{\mathfrak{p}}\left(x\right)=0.\label{eq:Momentum0}
\end{equation}
 The explicit normalized solution for this equation is found to be
\begin{equation}
\Phi_{\mathfrak{p}}\left(x\right)=\frac{\sqrt{\frac{\sqrt{\alpha^{2}+\beta^{2}}}{\pi}}}{\sqrt{1+\alpha^{2}x^{2}+i2\beta x}}\exp\left[-\frac{i\mathfrak{p}\arctan\left(\frac{\alpha^{2}x+i\beta}{\sqrt{\alpha^{2}+\beta^{2}}}\right)}{\hbar\sqrt{\alpha^{2}+\beta^{2}}}\right]\label{eq:Momentum1}
\end{equation}
in which $\mathfrak{p}$ is the continuous real eigenvalue of the
momentum operator $p$. To normalize the momentum eigenfunction (\ref{eq:Momentum1})
we introduce $x+i\frac{\beta}{\alpha^{2}}=\xi$ which yields
\begin{equation}
\Phi_{\mathfrak{p}}\left(\xi\right)=\frac{\sqrt{\frac{\sqrt{\alpha^{2}+\beta^{2}}}{\pi}}}{\sqrt{1+\frac{\beta^{2}}{\alpha^{2}}+\alpha^{2}\xi^{2}}}\exp\left[-\frac{i\mathfrak{p}\arctan\left(\frac{\alpha^{2}\xi}{\sqrt{\alpha^{2}+\beta^{2}}}\right)}{\hbar\sqrt{\alpha^{2}+\beta^{2}}}\right].\label{eq:Momentum2}
\end{equation}
On $\xi-axis$ (\ref{eq:Momentum2}) is normalizable in the usual
manner, i.e., by assuming $\xi$ to be a real variable it satisfies
\begin{equation}
\int_{-\infty}^{\infty}\left|\Phi_{\mathfrak{p}}\left(\xi\right)\right|^{2}d\xi=1.\label{eq:Momentum3}
\end{equation}
 Next, the corresponding $\mathcal{PT}$-symmetric Hamiltonian is
obtained to be as given in Eq. (\ref{eq:Hamil}). Upon applying the
PCT, introduced in Eqs. (\ref{GEMO2}) and (\ref{GEMO3}), the corresponding
Schrodinger equation (\ref{SE}) in $x$-space is transformed into
$z$-space given in Eq. (\ref{GEMO4}) in which the transformed coordinate
is found to be

\begin{equation}
z\left(x\right)=\frac{1}{\sqrt{\alpha^{2}+\beta^{2}}}\arctan\left(\frac{\alpha^{2}x+i\beta}{\sqrt{\alpha^{2}+\beta^{2}}}\right).\label{eq:9}
\end{equation}
We note that, $z\left(x\right)=\zeta+i\eta$ is a complex variable
such that
\begin{equation}
\tan\left(\omega\left(\zeta+i\eta\right)\right)=\frac{\alpha^{2}}{\omega}x+i\frac{\beta}{\omega}
\end{equation}
in which $\omega=$$\sqrt{\alpha^{2}+\beta^{2}}$. Eq. (\ref{eq:Momentum0})
implies
\begin{equation}
\frac{\alpha^{2}}{\omega}x=\frac{\sin\left(\omega\zeta\right)\cos\left(\omega\zeta\right)}{\cos^{2}\left(\omega\zeta\right)\cosh^{2}\left(\omega\eta\right)+\sin^{2}\left(\omega\zeta\right)\sinh^{2}\left(\omega\eta\right)}
\end{equation}
and
\begin{equation}
\frac{\beta}{\omega}=\frac{\sinh\left(\omega\eta\right)\cosh\left(\omega\eta\right)}{\cos^{2}\left(\omega\zeta\right)\cosh^{2}\left(\omega\eta\right)+\sin^{2}\left(\omega\zeta\right)\sinh^{2}\left(\omega\eta\right)}.
\end{equation}
The domain of $x$ is $\mathbb{R},$ such that (\ref{eq:Momentum1})
and (\ref{eq:Momentum2}) are a map of the form $\mathbb{R\rightarrow\mathbb{C}}.$
For a finite nonzero $\beta,$ at the limits $x\rightarrow\pm\infty,$
we find $\omega\zeta\rightarrow\pm\frac{\pi}{2}$ and $\omega\eta\rightarrow0.$ 

In Eq. (\ref{GEMO4}) we set the external potential to be zero, i.e.,
$V\left(x\right)=0$ for the entire domain of $x$ i.e., $x\in\mathbb{R}$.
Imposing $V\left(x\right)=0$ in (\ref{GEMO4}), one finds 
\begin{equation}
\frac{-\hbar^{2}}{2m}\frac{d^{2}}{dz^{2}}\chi\left(z\right)=E\chi\left(z\right)
\end{equation}
where $\chi\left(z\right)\rightarrow0$ when $\zeta\rightarrow\pm\frac{\pi}{2}$
and $\eta\rightarrow0.$ Hence, one writes the general solution for
(\ref{eq:Momentum3}) as
\[
\chi\left(z\right)=C_{1}\sin\left(Kz\right)+C_{2}\cos\left(Kz\right)
\]
in which $K^{2}=\frac{2mE}{\hbar^{2}}.$ Applying the boundary conditions,
one obtains the eigenfunctions and eigenvalues given by
\begin{equation}
\chi_{n}\left(z\right)=\left\{ \begin{array}{cc}
C_{2}\cos\left(n\sqrt{\alpha^{2}+\beta^{2}}z\right), & odd-n\\
C_{1}\sin\left(n\sqrt{\alpha^{2}+\beta^{2}}z\right), & even-n
\end{array}\right.
\end{equation}
and 
\begin{equation}
E_{n}=\frac{n^{2}\hbar^{2}\left(\alpha^{2}+\beta^{2}\right)}{2m},\label{eq:25}
\end{equation}
respectively.

\begin{figure}
\caption{\protect\includegraphics[scale=0.6]{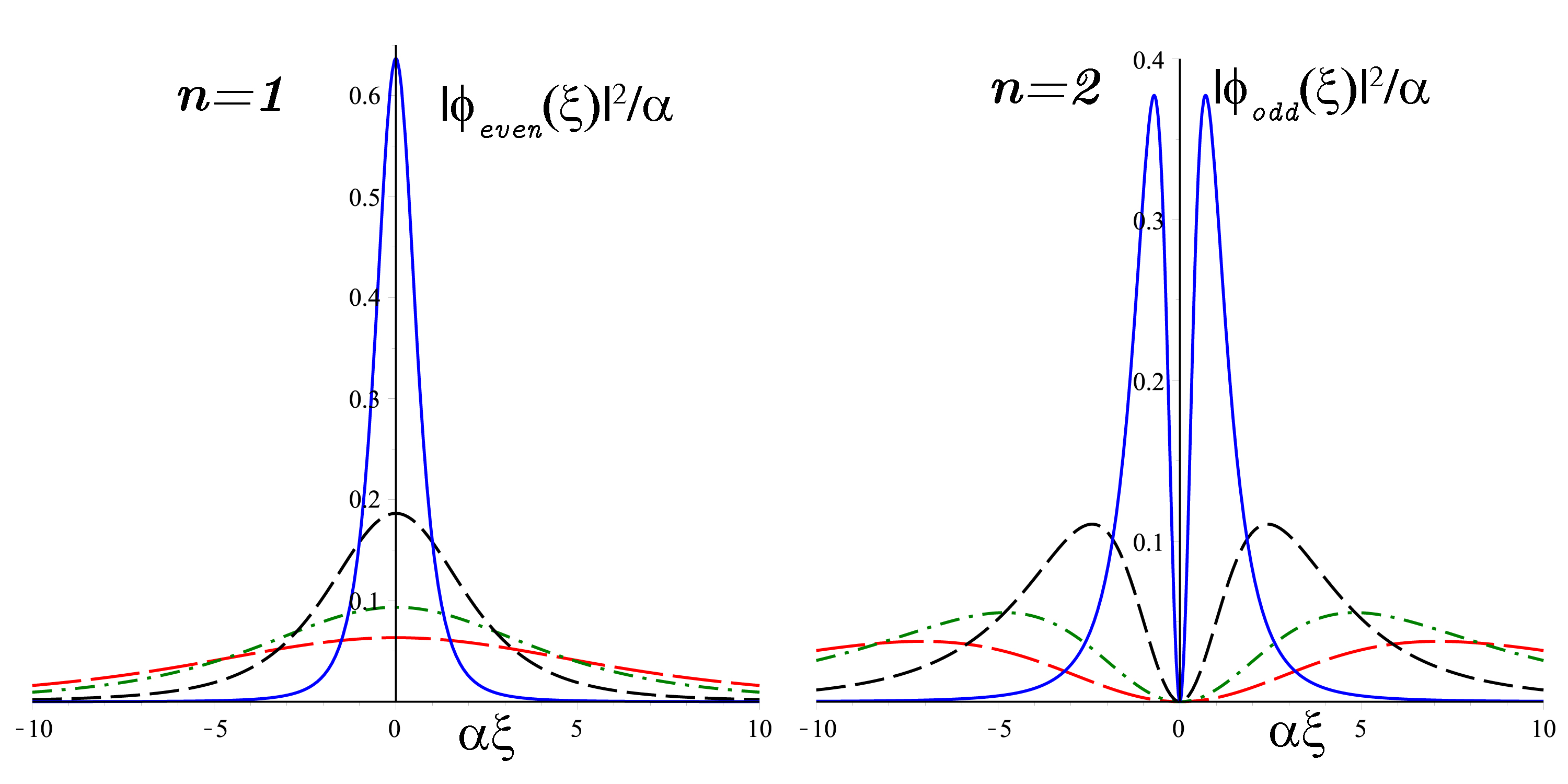}}

Plots of $\frac{1}{\alpha}\left|\phi_{n}\left(\xi\right)\right|^{2}$
in terms of shifted coordinate $\alpha\xi$ in which $\xi=x+i\frac{\beta}{\alpha^{2}}$
for $\frac{\beta}{\alpha}=0.0$ (blue, solid), $0.25$ (black, dash),
$0.50$ (green, dash-dot) and $1.0$ (red, long-dash) respectively.
With a given value for $\alpha,$ the effect of the parameter $\frac{\beta}{\alpha}$
is to decrease the confinement of the particle on $\xi$-axis. The
quantum number $n=1$ and $2$ for the left and right panels respectively.
\end{figure}

Please note that, $C_{1}$ and $C_{2}$ are the normalization constants
to be found. To find the corresponding wavefunction in $x$-space,
we write
\begin{equation}
\phi_{n}\left(x\right)=\frac{1}{\sqrt{1+\alpha^{2}x^{2}+2i\beta x}}\chi_{n}\left(z\right)\label{eq:10}
\end{equation}
which yields 
\begin{equation}
\phi_{n}\left(x\right)=\left\{ \begin{array}{cc}
C_{2}\frac{\cos\left(n\arctan\left(\frac{\alpha^{2}x+i\beta}{\sqrt{\alpha^{2}+\beta^{2}}}\right)\right)}{\sqrt{1+\alpha^{2}x^{2}+2i\beta x}}, & odd-n\\
C_{1}\frac{\sin\left(n\arctan\left(\frac{\alpha^{2}x+i\beta}{\sqrt{\alpha^{2}+\beta^{2}}}\right)\right)}{\sqrt{1+\alpha^{2}x^{2}+2i\beta x}}, & even-n
\end{array}\right.\label{eq:14}
\end{equation}
with real eigenvalues given in (\ref{eq:25}). Unlike $\chi_{n}\left(z\right),$
$\phi_{n}\left(x\right)$ can be normalized on the specific contour
$\mathfrak{C}$ such that
\begin{equation}
\int_{\mathfrak{C}}\left(\mathcal{\mathbf{\mathcal{CPT}}}\phi_{n}\left(x\right)\right)\phi_{n}\left(x\right)dx=\int_{\mathfrak{C}}\left(\phi_{n}\left(x\right)\right)^{2}dx\label{eq:15}
\end{equation}
in which $\mathcal{\mathbf{\mathcal{C}}}$ operator is known to be
the charge operator having eigenvalues $\pm1$ and commutes with $\mathcal{PT}$
operator \citep{Making sense}. As one can see in (\ref{eq:14}),
$\left(\phi_{n}\left(x\right)\right)^{2}$ is complex function implying
that it can not be the probability density. However, on the specific
contour $\mathfrak{C}$ on the complex $x$-plane where $\mathfrak{C}$
satisfies the certain conditions, i) $Im\left[\left(\phi_{n}\left(x\right)\right)^{2}dx\right]=0,$
ii) $Re\left[\left(\phi_{n}\left(x\right)\right)^{2}dx\right]\geq0$
and iii) $\int_{\mathfrak{C}}\left(\phi_{n}\left(x\right)\right)^{2}dx=1$,
it can be considered as the probability density \citep{Complex correspondence principle}.
Here, the contour $\mathfrak{C}$ is a line parallel to the real axis
defined by $Im\left(x\right)=\beta$ and the normalization constants
are found to be
\begin{equation}
C_{1}=C_{2}=\sqrt{\frac{2\sqrt{\alpha^{2}+\beta^{2}}}{\pi}}.
\end{equation}

In Fig. 1 we plot the ground state and the first excited state corresponding
to $n=1$ and $n=2$ in the solution (\ref{eq:14}). Increasing the
value of $\beta$ for a given $\alpha,$the probability density admits
a larger uncertainty for $\xi=x+i\frac{\beta}{\alpha^{2}}$. 

\section{Conclusion}

A $\mathcal{PT}$-symmetric quantum Hamiltonian admits a real energy
spectrum. For such a Hamiltonian, a complex potential is usually responsible
for the non-Hermiticity. Here in this research, we have shown that,
in the context of the GEMO one may introduce a non-Hermitian Hamiltonian
through a non-Hermitian GEMO. In doing so, we have introduced a $\mathcal{PT}$-symmetric
GEMO in virtu of a $\mathcal{PT}$-symmetric auxiliary function $\mu\left(x\right).$
The momentum eigenvalues are real and the Hamiltonian is non-Hermitian
but $\mathcal{PT}$-symmetric. Hence, the energy spectrum is real.
We have also presented an explicit example with a specific choice
for $\mu\left(x\right)$ for a quasi-free particle.


\begin{thebibliography}{1}
\bibitem{BenderBoettcher1998} C. M. Bender and S. Boettcher, Phys.
Rev. Lett. \textbf{80}, 5243 (1998);

C. M. Bender, S. Boettcher and P. N. Meisinger, J. Math. Phys. \textbf{40},
2201(1999);

P. Dorey, C. Dunning and R. Tateo, J. Phys. A: Math. Gen. \textbf{34},
391 (2001);

P. Dorey, C. Dunning and R. Tateo, J. Phys. A: Math. Gen. \textbf{34},
5679 (2001); 

C. M. Bender, M. V. Berry and A. Mandilara, J. Phys. A: Math. Gen.
\textbf{35}, 467 (2002).
\bibitem{Making sense}C. M. Bender, Rep. Prog, Phys. \textbf{70},
947 (2007).

\bibitem{Potentials}C. M. Bender and H. F. Jones, Phys. Rev. A \textbf{84},
032103 (2011);

C. M. Bender, M. DeKieviet and S. P. Klevansky, Philos. Trans. R.
Soc. A \textbf{371}, 20120523 (2013);

M. Znojil, Phys. Lett. A \textbf{264}, 108 (1999);

G. Levai, M. Znojil, Mod. Phys. Lett. A \textbf{16}, 1973 (2001); 

O. Mustafa, M. Znojil J. Phys. A: Math. Gen. \textbf{35}, 8929 (2002);

B. Bagchi, F. Cannata and C. Quesne, Phys. Lett. A \textbf{269}, 79
(2000).

K. G. Makris, R. El-Ganainy, D. N. Christodoulides and Z. H. Musslimani,
Int. J. Theor. Phys. \textbf{50}, 1019 (2011).
\bibitem{Complex correspondence principle}C. M. Bender, D. W. Hook,
P. D. Meisinger and Q. H. Wang, Phys. Rev. Lett. \textbf{104}, 061601
(2010).

\bibitem{M.H2}M. Izadparast and S. H. Mazharimousavi, Phys. Scrip.
(2020) (in press): arXiv:2002.10219.

\bibitem{Mostafazadeh}A. Mostafazadeh, J. Math. Phys. \textbf{43},
205 (2002);

A. Mostafazadeh, J. Math. Phys. \textbf{43}, 2814 (2002);

A. Mostafazadeh, J. Math. Phys. \textbf{43}, 3944 (2002); 

A. Mostafazadeh, J. Math. Phys. \textbf{44}, 974 (2003);

A. Mostafazadeh and A. Batal, J. Phys. A Math. Gen. \textbf{37}, 11645
(2004);

A. Mostafazadeh, Phys. Lett. B \textbf{650}, 208 (2007).
\bibitem{Experiments}A. Guo, G. Salamo, D. Duchesne, R. Morandotti,
M. Volatier-Ravat, V. Aimez, G. Siviloglou, and D. Christodoulides,
Phys. Rev. Lett. \textbf{103}, 093902 (2009);

C. Rüter, K. Makris, R. El-Ganainy, D. Christodoulides, M. Segev,
and D. Kip, Nat. Phys. \textbf{6}, 192 (2010);

K. F. Zhao, M. Schaden and Z. Wu, Phys. Rev. A, \textbf{81}, 042903
(2010);

S. Bittner, B. Dietz, U. Günther, H. L. Harney, M. Miski-Oglu, A.
Richter and F. Schäfer, Phys. Rev. Lett. \textbf{108}, 024101 (2012).
\bibitem{M.H1}M. Izadparast and S. H. Mazharimousavi, Phys. Scr.
\textbf{94}, 115215 (2019).

\bibitem{Costa Filho} R. N. Costa Filho, J. P. Braga, J. H. Lira
and J. S. Andrade Jr, Phys. Lett. B \textbf{755}, 367 (2016).
\end{thebibliography}
\end{document}